# 人-AI 交互：实现“以人为中心 AI”理念的跨学科新领域

许为[1]，葛列众[1]，高在峰[2]

（1. 浙江大学 心理科学研究中心 杭州 310058；2.浙江大学 心理与行为科学系 杭州 310058）

**摘 要：** AI技术造福了人类，也给研发带来了挑战，如果开发不当，会伤害人类和社会。目前国内外还没有系统的跨学科工作框架来有效地应对这些新挑战。为顺应学科发展的交叉趋势，中国国家自然科学基金委2020年成立了交叉科学部。在这样的背景下，本文分析AI系统研发面临的新挑战，进一步阐述我们在2019年提出的“以人为中心AI”（HCAI）研发理念和设计目标。HCAI研发理念在国外目前是AI界的热门课题之一，为推动 HCAI 理念的落实，我们系统地提出了人-人工智能交互（HAII）的跨学科新领域，定义了其目的、范围、研究和应用重点等。通过文献综述和分析，本文总结了国内外HAII研究和应用的重点，提出了今后的主要研究方向。最后，针对今后HCAI理念和HAII领域的工作，提出了一系列对策和建议。

# Human-AI interaction:

# An emerging interdisciplinary domain for enabling human-centered AI

XU Wei [1], GE Liezhong [1], GAO Zaifeng [2]

(1. Zhejiang University, Center for Psychological Sciences, Hangzhou 310058, China;

2. Zhejiang University, Department of Psychology, Hangzhou 310058, China)

**Abstract** The new characteristics of AI technology have brought new challenges to the research and development of AI systems. AI technology has benefited humans, but if improperly developed, it will harm humans. At present, there is no systematic interdisciplinary approach to effectively deal with these new challenges. This paper analyzes the new challenges faced by AI systems and further elaborates the “Human-Centered AI” (HCAI) approach we proposed in 2019. In order to enable the implementation of the HCAI approach, we systematically propose an emerging interdisciplinary domain of "Human-AI Interaction" (HAII), and define the objective, methodology, and scope. Based on literature review and analyses, this paper summarizes the main areas of the HAII research and application as well as puts forward the future research agenda for HAII. Finally, the paper provides strategic recommendations for future implementation of the HCAI approach and HAII work.

**通信作者**：许为，xuwei11@zju.edu.cn

# 1 引言

人工智能（AI）技术正在造福人类，但是，目前许多AI系统的研发主要遵循“以技术为中心”的理念[1-5]。研究表明不恰当的AI技术开发导致了许多伤害人类的事故,AI事故数据库已经收集了1000多起事故[6],这些事故包括自动驾驶汽车撞死行人,交易算法错误导致市场“闪崩”,面部识别系统导致无辜者被捕等。美国工程院院士、计算机教授Shneiderman[4]将围绕“以技术为中心”还是“以人为中心”理念开发 AI 系统的争议形象化地描述为“AI哥白尼革命”,提出AI开发应该将人类放在中心,而不是算法和AI技术。

近几年来，围绕“以人为中心AI”理念、如何避免 AI 伤害人类以及产生社会负面影响等方面的研究引起越来越多的重视[2-3][5-11],目前国内外还没有形成系统化的跨学科工作框架来有效应对这些新挑战及促进这方面工作的开展。中国国家自然科学基金委员会在 2020 年成立了交叉学科部，在交叉科学高端学术论坛上，受邀的 AI、人机交互（human-computer interaction)、人因工程(human factors engineering）等专家一致认为，学科交叉是未来科学发展的必然趋势。

在这样的基于跨学科合作理念的背景下，本文回答以下三个问题： 与传统计算技术相比， AI技术带来了什么新挑战？应该如何促进“以人为中心AI”理念在 AI 研发中的应用？从跨学科合作角度我们应该采取什么策略？本文将进一步阐述我们在2019年提出的“以人为中心AI”(human-centered AI，HCAI）理念[2]，系统地提出人-人工智能交互（human-AI interaction，HAII）这一新兴跨学科领域。希望通过倡导HCAI理念和 HAII领域，促进AI研发造福于人类，避免潜在的负面影响。

# 2 AI 技术带来的新变化和新挑战

## 2.1 AI技术的跨时代特征

AI界一般认为AI技术主要经历了三次浪潮。前两次浪潮集中在科学探索,局限于“以技术为中心”的视野，呈现出“学术主导”的特征。深度机器学习、算力、大数据等技术推动了第三次浪潮的兴起。在第三次浪潮中，人们开始重视AI技术的应用落地场景，开发对人类有用的前端应用和人机交互技术，考虑AI伦理等问题。同时，AI界开始提倡将人与AI视为一个人机系统,引入人的作用[1][3]。

可见，第三次浪潮开始围绕“人的因素”来开发AI，促使人们更多地考虑“以人为中心AI”的理念。因此，第三次浪潮呈现出“技术提升 + 应用开发 + 以人为中心”的特征[2]，意味着AI开发不仅是一个技术方案，还是跨学科合作的系统工程。

## 2.2 智能时代的新型人机关系

AI可以开发成具有自主化（autonomy）特征的智能体。取决于自主化程度，AI系统可以拥有一定程度上的类似于人的认知、学习、自适应、独立执行操作等能力，在特定的场景下可以自主地完成一些特定任务，可以在一些设计未预期的场景中自主地完成以往自动化技术所不能完成的任务[9-11]。

这种智能自主化特征赋予人机系统中机器新的角色。在非智能时代，人类操作基于计算技术，机器充当辅助工具角色。人与 AI 系统的交互本质上是人与自主智能体的交互。随着AI技术提升,自主智能体有可能从一种支持人类操作的辅助工具的角色发展成为与人类操作员共同合作的队友，扮演“辅助工具 + 人机合作队友”的双重新角色[11]。

因此，智能时代的人机关系正在演变成为团队队友关系，形成一种“人机组队”（human-machine teaming）式合作[13-14]。 智能时代的这种人机关系区别于PC时代的人机交互，对AI研发是挑战和机遇,研发者需要在AI研发中要利用这种人机合作，保证人类能够有效控制AI系统，避免伤害人类。

## 2.3 人-非AI系统交互与人-AI系统交互的比较

人机交互是PC时代形成的跨学科领域,它研究人-非AI计算系统之间的交互。表1比较了人-非AI系统交互与人-AI系统交互之间的一些特征。人-AI系统交互所具备的特征是基于AI系统具有较高的智能自主化程度,有些特征目前还没实现。从表1可见，与人-非智能系统交互相比，人-智能系统交互带来了许多新特征和新问题,也给人-智能系统交互的研究和应用带来了新机遇。

在人-非智能系统交互中，作为一种支持人类操作的辅助工具，机器依赖于事先设计的规则和算法。尽管人机之间也存在一定程度上的人机合作，但是作为辅助工具的机器是被动的，只有人可以主动地启动这种有限的合作。

AI 系统智能体具备的自主化特征使得智能体与人类之间可以实现一定程度上类似于人-人团队之间的“合作式交互”。在特定的操作环境中，这种交互可以是由两者之间双向主动的、分享的、互

补的、可替换的、自适应的、目标驱动的以及可预测的等特征所决定的（见表 1）。随着 AI 技术的发展，未来 AI 系统将更多点地具备这些特征[12]。

由此可见，智能时代人-AI 系统交互的新特征以及研究的问题等已经远超出了目前人机交互研究和应用的范围，需要一种新思维来考虑如何更加有效地开展多学科合作来应对人-AI 系统交互以及 AI 系统研发中面临的一系列新特征和新挑战。

## 3 HAII 领域的兴起及领域理念

国外针对人-AI 系统交互的研究和应用已经展开[13-14]。例如：人-智能体交互[15]，人-自主化交互(human-autonomy interaction)[16],人-AI 交互[17]。尽管这些工作各有侧重点，但是都是研究人与智能“机器”(智能体、智能代理等)之间的交互。所以，这种交互本质上就是人-AI 交互 （human-AI interaction，简称 HAII)。目前还没有一个系统的有关HAII领域的工作框架,有必要正式提倡将HAII作为一个新的多学科交叉领域来推动。

### 3.1 HAII领域的理念：以人为中心AI

近几年,当“以技术为中心”方法影响着 AI 研发的同时，研究者也在探索基于“以人为中心”的 AI 开发方法，例如，以人为中心的算法,AI 人文设计，包容性设计，基于社会责任的 AI[8]。

斯坦福大学在 2019 年成立了“以人为中心 AI”

**表 1 人-非 AI 计算系统交互与人-AI 系统交互的特征比较**

**Table 1 Comparative analysis between human interaction with non-AI systems and AI systems**

| 特征 | 计算机时代的人-非 AI 计算系统交互 | 智能时代的人-AI 系统交互 |
|---|---|---|
| 实例 | 办公软件，洗衣机，自动生产线等 | 智能音响，智能决策系统，自动驾驶汽车等 |
| 机器行为和智能 | 按照固定算法、逻辑和规则产生确定的机器行为;不具备机器智能 | 具有不同程度的类似于人的认知能力（学习、自适应、自我执行等）；展示特殊、可演化的机器行为 |
| 机器角色 | 主要作为一种辅助工具 | 也可能成为与人类合作的团队队友 |
| 机器输出 | 具确定性 | 具不确定性 |
| 人类操作员角色 | 监视员，执行者 | 也可能成为与 AI 合作的队友（人应是最终决策者） |
| 人机关系 | 人机交互 | 人机交互 + 人机组队式合作 |
| 用户界面 | 图形用户界面，触摸屏交互，显式交互等 | 也包括新型智能交互：语音交互，人脸识别，脑机界面，隐式交互等 |
| 人机交互的行为特征 | 由人启动的、基于显式界面的人机交互 | 基于人的认知、行为、情感、场景上下文等信息，智能体也可以主动启动基于隐式界面的人机交互等 |
| 启动能力 | 人主动启动任务、行动，机器被动接受 | 人机双方均可主动地启动任务、行动 |
| 人机交互的方向性 | 只有人针对机器的单向式信任、情景意识、决策等 | 人机之间双向式的信任、情景意识、意图，人机之间可分享的决策控制权（人应拥有最终控制权） |
| 智能互补性 | 机器无智能，人与机器之间无智能互补 | 机器智能与人的生物智能之间的互补 |
| 系统输出的可解释性 | 主要取决于系统输出界面的可用性 | 还呈现 AI“黑匣子”效应，导致系统输出难以解释和理解 |
| 人、机器的预测能力 | 仅人类操作员拥有 | 人、机器均可借助行为、情景意识等模型，预测对方的行为、环境、系统等状态 |
| 自适应能力 | 仅人类操作员拥有 | 人、机器均可适应对方的行为及操作场景 |
| 目标设置能力 | 仅人类操作员拥有 | 人、机器均可设置或调整系统目标 |
| 替换能力 | 机器可以替换人的任务（借助于自动化技术，主要是体力方面） | 机器可以替换人的体力、认知任务（人机之间可主动或被动地接管、委派任务等） |
| 人机合作 | 有限 | 基于以上一些特征，可能产生更有效的人机合作 |
| 用户需求 | 主要包括可用性、心理、安全、生理等 | 还包括情感、隐私、伦理、决策自主权等 |

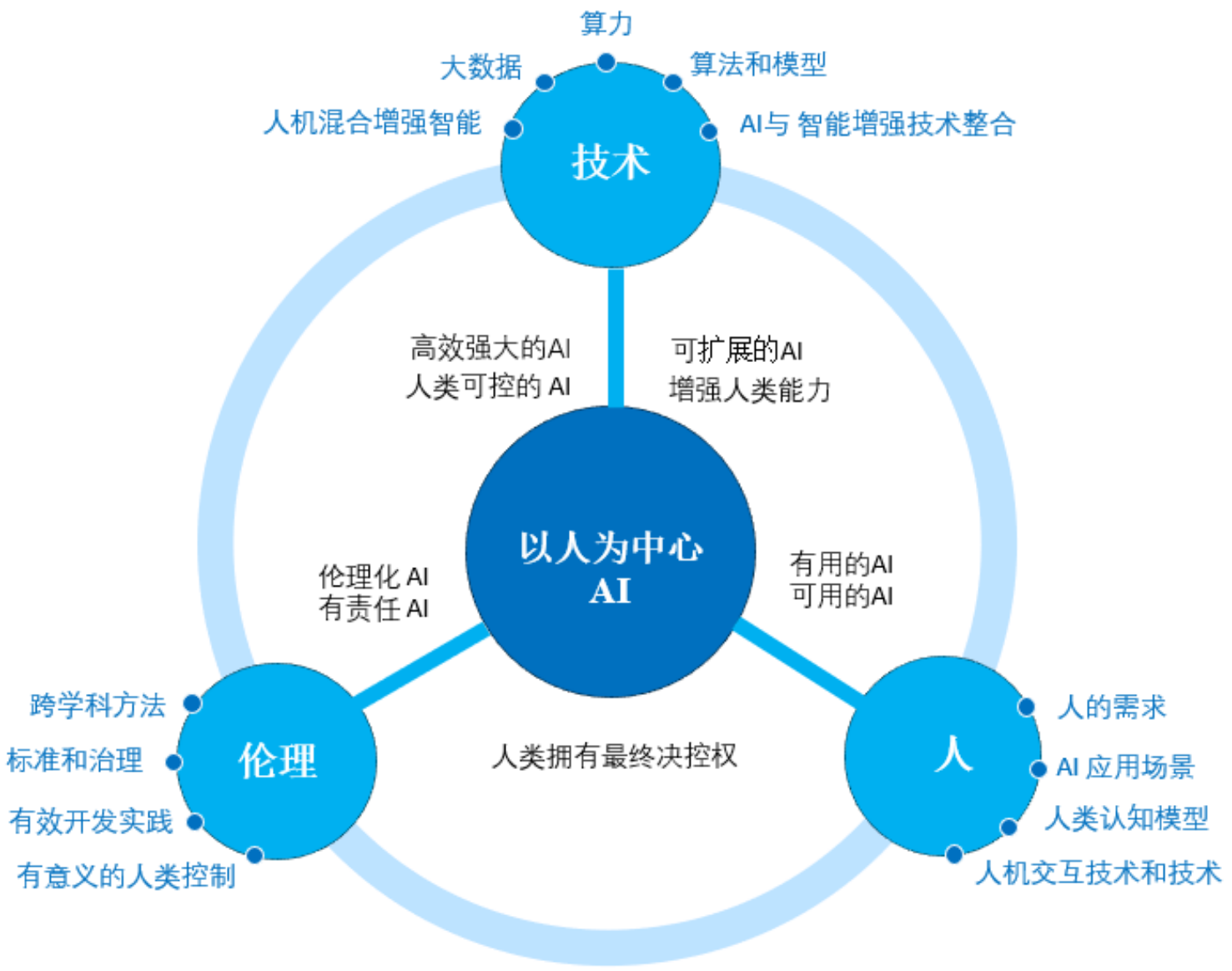

图1 以人为中心 AI（HCAI)理念（修改自：Xu，2019）

Figure 1 Human-Centered AI (HCAI) design philosophy (adapted from, Xu, 2019)

(Human-Centered AI,以下简称 HCAI) 研究中心，目的是通过技术提升与伦理化设计手段，开发出合乎人类道德伦理和惠及人类的 AI 系统[18]。

许为[2][19]在 2019 年提出了一个“以人为中心 AI”(HCAI)的系统概念框架，该框架包括人、伦理、技术三个方面。Shneiderman[20]在 2020 年提出了一个为开发可靠、安全和可信赖的 AI 系统的指导框架。HCAI 就是指导 HAII 新领域的理念。我们以下进一步阐述 HCAI 理念的三个方面： 技术、人、伦理(见图 1)。其中,图 1 概括了各方面工作的主要途径（见图 1 中围绕三个周边圆形部分的蓝色字体），例如，人的需求，AI 应用场景；图 1 也概括了这些工作要达到的 HCAI 设计目标（见图 1 中围绕“以人为中心 AI”中心圆形部分的黑色字体），例如，可用的 AI,有用的 AI。

（1）“技术”方面：强调 3 个部份的有机结合。（a）机器智能: 利用算法、大数据、算力等技术来开发机器智能;（b）人类智能：利用智能增强技术，借助心理学、脑神经技术等方法推动人类智能的增强（见 4.1）；(c)人机混合增强智能：AI 界已经认识到单独发展 AI 技术的路径遇到了瓶颈效应，在高级人类认知方面难以达到人类的智能水平[1][3][20]。 因此,HCAI 理念强调将人的作用融入人机系统,通过人机智能的互补,开发人机混合增强智能、AI 与人类智能增强技术的整合（见 4.1、4.2)。目的是开发出可持续发展、强大、人类可控的 AI；AI 开发的目的是提升人的能力， 而不是取代人类。

（2）“人”方面：强调在 AI 系统研发中从人的需求出发，落实有效的应用场景， 开发人类认知模型，在 AI 研发中实施基于“以人为中心”的人机交互设计和方法(建模、设计、测试等)。目的是开发出有用的(满足人的需求、有使用价值)、可用的(易用、易学)、人类拥有最终决控权的 AI 系统。

（3）“伦理”方面：结合跨学科方法、有效开发实践、标准和治理等工作,通过工程设计手段(如“有意义的人类控制”，详见 4.7)，保证 AI 开发遵循公平、人的隐私、伦理道德、人的决策权等方面的权益。目的是开发出伦理化、负责任的 AI。

HCAI 理念强调在 AI 开发中保持人的中心地位，贯彻技术、人、伦理三方面相互依承的系统化 AI 开发思维， 主张 AI 开发是一个跨学科协作的系统工程， 开发出可靠的、安全的、可信赖的 AI 系统[2][20]。

## 3.2 HAII领域的工作框架

针对 HAII 这一新兴领域,我们作出以下初步的定义。图 2 示意了人-人工智能交互（HAII）的领域框架，其中，蓝色圆圈部份代表跨学科的主要合作学科，白色长形部份代表本文所讨论的 HAII 研究和应用的主要问题（详见 4.1 至 4.7)。

• HAII 领域理念：以人为中心 AI（HCAI)。

• HAII 领域目的：作为一个跨学科交叉领域，HAII 利用 AI、计算机科学、人机交互、人因工程、心理学等学科技术和方法，致力于合作研发 AI 系统，优化人与 AI 系统之间的交互，注重机器与人类智能的优势互补，全方位考虑 AI 伦理道德，强调人

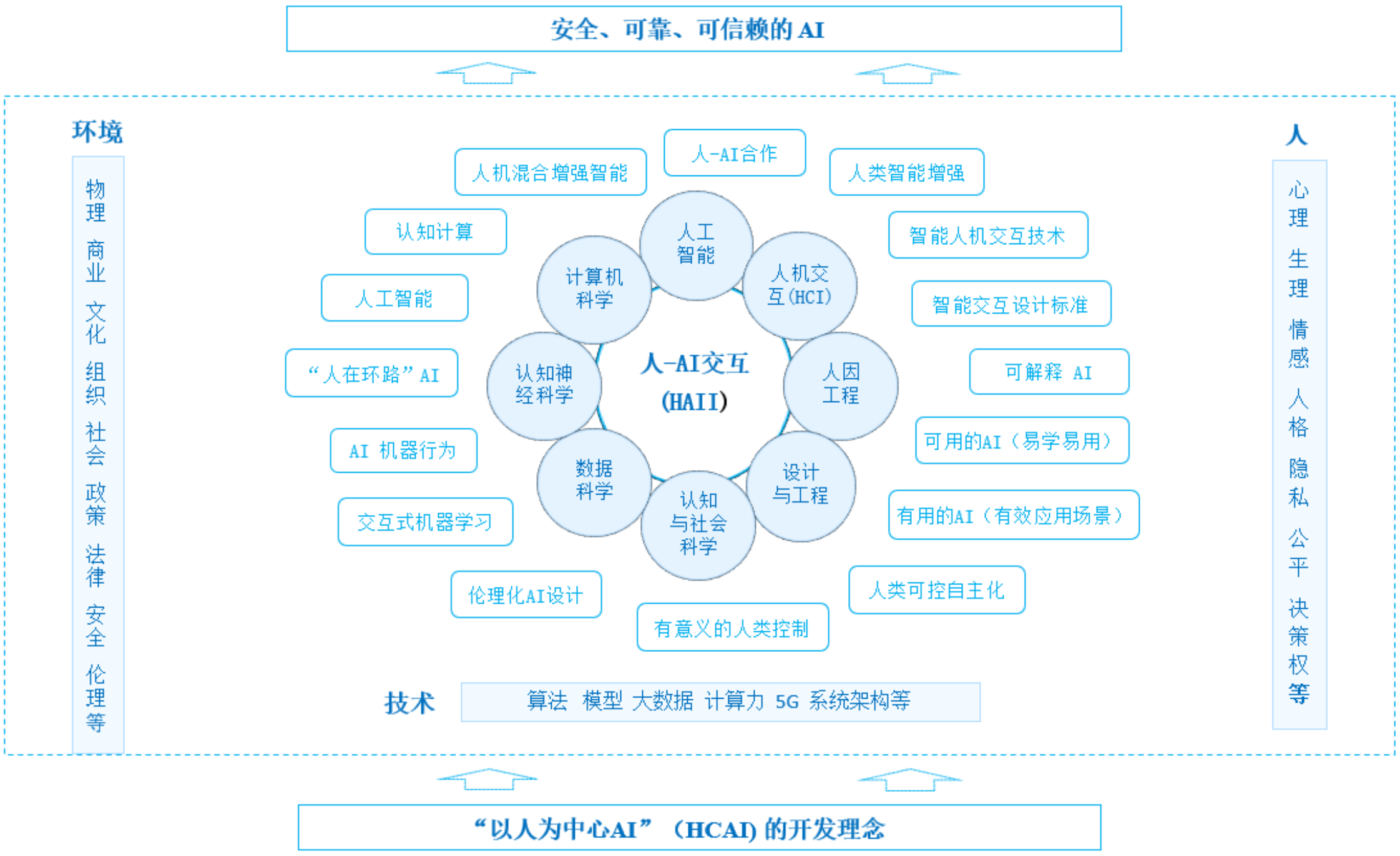

图2 人-人工智能交互（HAII）领域示意图

**Figure 2 Illustration of the Human-AI Interaction (HAII) domain**

对 AI 系统的最终决控权，通过提供一个跨学科的合作平台，在 AI 系统开发中实现 HCAI 开发理念，为人类提供安全、可靠、可信赖的 AI。

- HAII 研究和应用范围：狭义地说，HAII 涉及到人与 AI 系统交互的研究和应用；广义地说，任何涉及到由人来使用、影响人的 AI 研究发都属于 HAII 的范畴，包括与人产生交互的 AI 系统的研究和应用领域，如智能手机应用 APP，智能人机交互技术，智能决策系统，智能物联网等。如图 2 所示，HAII 从人-机-环境系统的角度来考虑各种因素对人与 AI 交互的影响，全面了解这些影响有助于发挥 AI 技术的优势，避免负面影响。
- HAII 领域方法：作为一个跨学科领域，通过多学科方法（建模、算法、设计、工程、测试等）和合作的流程来开发 AI 系统。这些方法来自这些相关学科，例如计算模型、工程设计方法、行为科学研究方法、人机交互设计等。
- HAII 领域人员：从事 HAII 研究和应用的人员包括来自 AI、计算机、数据科学、人机交互、心理学、认知神经科学、社会科学等专业人员。广义地来说，大多数从事 AI 系统研发的人员都属于这个范畴，他们研发的 AI 系统或多或少都与人交互，都是为了开发出有利于人类的 AI 系统。

## 3.3 为什么需要 HAII 新领域

首先，HAII 领域为各学科提供了一个合作平台。HAII 有助于在一个领域名称（HAII）下，联合参与 AI 系统研发的跨学科、跨行业专业人员，避免易混淆的名称，这种跨学科合作有助于有效地开发以人为中心的 AI 系统。

其次，HAII 领域强调其研究和应用的对象是 AI，不是传统的非 AI 系统，有助于提醒人们注重 AI 与非 AI 系统之间的特征差异，促使人们重视 AI 技术带来的新挑战和新问题，采用有效的方法来解决 AI 系统开发中的独特问题。

最后，HAII 领域有助于推动 HCAI 理念在 AI 研发中的落实。HAII 强调 AI 研发中将人的中心作用整合到系统设计中，避免潜在安全风险问题[8][12]。

历史上，新技术促进了新领域的产生。进入 PC 时代，传统“人机交互”（human-machine interaction）领域过渡到新版的“人机交互”（人-计算机交互），但是此“机”非彼“机”[21]。智能时代的机器过渡到 AI 系统，AI 的新特征促使 HAII 新领域的产生，因此，HAII 的出现也是必然的。

我们并不建议将 HAII 设置为一门独立的新学科，强调 HAII 是一个新型跨学科领域，希望通过该领域的跨学科、跨行业协同合作来落实 HCAI 理念。例如，HAII 工作需要人机交互人员的参与，他们必须采用新思维开展针对人与 AI 交互的研究和应用。

## 3.4 实现HCAI理念面临的挑战和HAII解决方案

HAII 领域的工作并非刚刚兴起，AI 界和其他

相关学科已经开展了一些工作。为进一步阐述 HAII 领域，基于文献综述和分析，表 2 概括了当前 HAII 研究和应用的重点、实现 HCAI 理念的挑战、HAII 领域可能的解决方案、期望的 HCAI 设计目标。表 2 也引用一些实例，详细内容在本文第 4 部分讨论。

从表 2 可知, 首先,为实现基于 HCAI 理念的设计目标，HAII 研究和应用有许多挑战期待解决。不解决这些挑战我们就无法实现 HCAI 理念，无法开发出安全、可靠、可信赖的 AI。

其次，表 2 轮廓出 HAII 研究和应用的范围。目前来自 AI 和其他学科的专业人员在开展这方面的工作，这些挑战不是单一学科可以解决的，这正说明了 HCAI 和 HAII 工作需要跨学科的协同合作。

最后，HAII 领域采用跨学科的方法。HAII 研究和应用的挑战、可能的解决方案以及实例都依赖于跨学科方法（建模，工程设计，行为科学方法等），单一学科的方法无法有效地解决这些问题。

40 多年前，当 PC 新技术刚兴起时，开发者基本遵循 “以技术为中心”的理念。 随着 PC 的普及，许多用户体验问题随之出现，人们开始意识到“以人为中心设计”理念的重要性，来自计算机科学、人因工程、心理学的专业人员协同推动了人机交互学科的形成和发展。多年的实践，用户体验的理念已经在社会和计算技术界形成共识。

今天，随着 AI 技术的引进，我们又目睹了类似情景，但是这一次忽略“以人为中心”理念的代价对人类和社会的影响将更为严重[6]。因此，各学科必须再一次协同合作，推动智能时代的“以人为中心设计”版本(即 HCAI 理念)和 HAII 领域的工作，更加有效地利用 AI 技术，扬长避短，为人类服务。

另外，HCAI 理念和 HAII 领域定义的是 AI 开发中应该遵循的理念、目标及途径等，并非是一个具体模型或算法。我们强调跨学科合作，一旦明确了这些理念、目标及途径等，AI 人员就能更加有效地开发出实现 HCAI 理念、设计目标的模型、算法以及技术。

# 4 HAII 研究和应用的关键问题

依据 HCAI 理念，针对目前 AI 带来的新挑战，我们从以下几方面分析目前 HAII 研究和应用的进展，提出今后的重点方向。目前一些 AI 人员也在开展这方面工作，希望 HCAI 理念和 HAII 领域的提出能够强化这些人员的 HCAI 理念以及跨学科的合作，也希望非 AI 人员积极参与到 HAII 研究和应用中。

## 4.1 人类智能增强

自 1956 年 AI 概念被提出后，研究者已经开始致力于另一条路径: 智能增强 (Intelligence Augmentation，IA)[22]。智能增强致力于增强人类智能[23-24]。研究人员利用新技术（如心理学、脑机接口、虚拟现实），借助 AI 技术来推动智能增强的研究和应用[52]。 从 HCAI 理念看，HAII 领域与智能增强具有相同目标：利用 AI 技术来增强人的能力。

AI 领域与智能增强领域之间长期存在竞争。一些 AI 人员认为 AI 可以取代人类，而智能增强人员认为 AI 仅仅为智能增强技术提供了新的手段[25]。从 HCAI 理念分析，AI 和智能增强会采用类似的技术，追循的应该都是扩展人类智能，应该是“以人为中心”的伙伴关系，许多智能方案其实是两种技术的集合，HAII 可以为两者的合作起到桥梁的作用。

智能增强研究中有许多问题今后需要 HAII 领域的贡献。首先，机器智能无法模仿人类智能的某些维度，HAII 提倡从跨学科角度来探索哪种类型的人类智能增强以及技术可以提供有效手段来弥补 AI 的弱点[26]，这需要智能增强人员主动寻求来自 AI、心理学、认知神经科学、人因工程学科的支持。

其次，开发 AI 与智能增强技术最佳组合的应用解决方案将有效促进两种技术之间的合作，从而达到“1 + 1 > 2“的效果。HAII 领域可以起到一个中间桥梁的作用，从人机交互、心理学等学科角度，从人机交互方式、多模态交互兼容性、人类认知加工水平、AI 系统自主化程度等多种维度来开发能够支持人与智能系统有效交互的解决方案。

第三，依据 HCAI 理念，智能增强专业人员要将人类置于系统方案的中心。HAII 领域提倡心理学、认知神经科学等人员积极参与研究。 例如，基于可塑性机制, 构建认知负荷可控、及时生理反馈、体脑双向交互的新型人机交互研究。这些研究将有效支持在许多应用领域中训练和增强人类智能[53]。

最后，在生物神经层面上寻找 AI 技术与智能增强技术的整合解决方案。这是当前关注的研究方向之一,例如，脑机融合[27]。HAII 提倡 AI、认知神经科学、脑成像技术、人机交互等人员合作，优化脑机界面解决方案，通过在生物神经层面上整合 AI 与智能增强技术来探索有效的人-AI 交互手段 [28]。

## 4.2 人机混合增强智能

将人类的作用和人类智能引入 AI 系统将形成人类智能与机器智能的优势互补，从而开发出更强大、可持续发展的人机混合增强智能[1][3][20][28]。

表 2　HAII 领域中实现 HCAI 理念所面临的挑战、HAII 可能解决方案、应用实例以及 HCAI 设计目标

Figure 2　Challenges in realizing the HCAI design philosophy, possible HAII solutions, application examples, and expected HCAI design goals

| HAII 领域重点（见第 4 部分） | 实现 HCAI 理念面临的挑战 | 基于 HCAI 理念的 HAII 可能解决方案 | HAII 领域研究和应用实例 | 基于 HCAI 理念的设计目标（图 1） |
|---|---|---|---|---|
| AI 系统的机器行为（第 4.7 节） | 潜在的、带偏见的系统输出，意外的机器行为，独特的机器行为演变，不成熟的机器学习训练和测试方法，缺乏用户参与的机器学习，社会交互中的复杂机器行为，多重 AI 代理之间的复杂行为和交互 | “以人为中心”机器学习，交互式机器学习，人机交互方法在数据收集、培训、算法调整、测试中的应用，基于行为科学方法的机器行为研究 | 交互式机器学习:[29] 基于“以人为中心”理念，由目标用户（领域专家）直接参与，降低对机器学习专家的依赖，通过有效人机交互来构建和训练机器学习模型。用户检查、训练模型结果，不断调整后续输入直到获得满意结果。相对于传统机器学习，该方法更加快速、高效和优化，已应用在推荐系统、信息检索、情景感知等领域 | 有用的 AI，伦理化 AI，有责任 AI |
| 人类智能增强（见第 4.1 节） | 人类智能增强技术（IA）与 AI 技术之间的竞争，IA 和 AI 技术之间缺乏优势互补的最佳方案 | HAII 可发挥 IA 和 AI 间的桥梁作用，制定两种技术的最佳组合方案，确保 AI 技术增强人类智能，保证人拥有最终控制权，开发生物神经层面的方案（脑机融合等） | 整合 IA 和 AI 技术的智能系统，在应用中增强人类智能，人类通过监控（远程等）实现人机协同合作以及保证人的决控权[25]。 例如，智能无人机（军用类等）、机器人（月球探测等危险场景、医疗外科精细手术等）、智能决策系统（股票交易系统等） | 有用的 AI，可用的 AI，增强人的能力，人拥有最终决控权 |
| 人机混增强合智能（见第 4.2 节） | 机器智能难以模拟人类高级认知能力，机器智能技术发展的瓶颈效应，孤立地开发机器智能的发展途径缺乏可持续性，AI 开发中缺乏对人类控制和安全性的充分考虑 | 人机混合增强智能，“人在环路”AI 系统及交互设计，人机协同控制，脑机混合系统，基于认知心理学研究的认知计算（情感、意图等），人类高级认知能力模型、知识表征和图谱，人机共生与融合 | “人在环路”混合增强智能 [1][30-31]：利用人与机器智能的优势互补，处理大规模、不完整和非结构化知识信息；用户与 AI 系统交互中不断知识迭代和学习，加深对数据及系统理解，AI 模型接受特定输入并根据用户反馈信息确定输出，达到优于各自单独实现的结果，避免 AI 技术带来的失控风险，已应用在自动驾驶、辅助医疗、视频检索等领域 | 可扩展的 AI，高效强大的 AI，人类可控的 AI，人拥有最终决控权 |
| 人- AI 合作（见第 4.3 节） | 缺乏成熟的人- AI 合作理论、方法、认知架构，缺乏成熟的人机态势感知共享、人机共信、人机心理模型共享、人机决策共享的理论、模型和方法 | 人-AI 合作理论和模型，人-AI 合作团队绩效评估和测试方法，人机共驾，社会环境中的人-AI 合作，人-AI 合作的人机交互模型，人作为最终决策者的人-AI 合作设计 | 从多学科角度出发，从感知、认知、执行层面上为基于人-AI 合作的 AI 解决方案提供依据。 例如，人-AI 合作研究策略和框架[32]，人-AI 合作团队绩效评估[33]， 人-AI 互信[34]，人-AI 合作中的心理模型[35]，人-AI 合作的系统设计[36]，人-AI 合作系统权限[37]，人-AI 合作的定量和定性建模[38] | 有用的 AI，可用的 AI，人类可控的 AI |
| 可解释的 AI（见第 4.4 节） | AI“黑匣子”效应，用户无法理解 AI 系统决策，影响人类决策，影响 AI | “以人为中心”的可解释 AI，终端用户参与式的可解释 AI，可理解 | “以人为中心”的实时化可解释 AI 方案：在[39]的计算机游戏研究中，AI 代理与用户交互中实时生成基于自然语言 | 可用的 AI，有责任 AI |
| | 技术推广，心理学解释理论没得到应用，没有终端用户参与的方法， AI 是可解释的但是终端用户无法理解 | AI， AI 系统输出界面可视化设计，“人在环路”式可解释 AI，心理学解释理论的转化应用，用户参与式、交互式 AI 人机界面设计 | 的推理数据，使用这些数据训练 AI 模型，该模型能够对游戏结果生成人类可理解的推理。在[40] 的研究中，自动驾驶汽车乘客通过有效的人机交互手段选择行车环境目标，这些选择目标使驾驶算法的决策更容易解释和理解 | |
| 人类可控自主化（见第 4.5 节） | AI 独特的自主化特征以及潜在的负面影响，对自动化与自主化概念的混淆，低估了的自主化影响（被混淆为高水平自动化） | 人类可控自主化，人机共享自主化，人-自主化组队合作，自动化研究成果转化，有意义的人类控制，自主化故障追踪数据系统 | Udelv 自动送货车（ADV）[41]，ADV 采用 L4 自动驾驶，可以在一些公共场所范围内完成无人驾驶点对点操作。该系统包括远程监控系统，如果需要时（任务边缘区、应急场景等）， 通过无缝式人机操控转换可以实现人工干预操纵 | 可用的 AI，<br>伦理化 AI，<br>人类可控的 AI，<br>人拥有最终决控权 |
| 智能人机交互（见第 4.6 节） | 缺乏针对智能交互的人机交互范式，复杂智能计算环境中人类有限认知资源的瓶颈效应，针对非 AI 系统的现有人机交互设计标准，智能交互的用户体验问题 | 智能人机交互新范式，有效的智能人机交互设计，针对 AI 系统的人机交互设计标准，基于人-AI 合作的人机界面设计，智能人机交互的可用性设计 | 针对 AI 系统的人机交互设计标准： ISO 人-AI 系统交互的技术文件（ISO 9241-810）[42]，微软设计准则和指南[17]，“Google AI + People Guidebook”[43]；有效的智能人机交互开发流程：“配对式 AI 合作开发流程”[44]；有效的人机交互设计方法：AI 优先方法[43]，AI 作为设计材料[45] | 可用的 AI，<br>有用的 AI |
| 伦理化 AI 设计（见第 4.7 节） | 人类可能缺乏对 AI 系统的最终控制权，AI 系统产生输出偏差和意外结果，滥用 AI 系统（导致歧视，隐私泄密等），缺乏对 AI 系统故障的追溯和问责机制 | 有意义的人类控制，AI 错误追溯机制，透明化设计，优化机器学习建模、训练、测试，AI 人员知识提升，跨学科方法的 AI 伦理化设计，伦理化设计技术和示例 | “有意义的人类控制”设计[46]：透明化系统设计、有效的人机交互，人类操作员能够对所用的自主技术拥有足够的信息（态势感知等）来确保做出知情且有意识、合法的决策；装备“故障追踪系统”机制，实现系统行为和故障的追溯和问责制 | 伦理化 AI，<br>负责任 AI，<br>可用的 AI，<br>人类可控的 AI |

目前，针对混合增强智能的研究基本可分为两类。 第一类是在系统层面上的“人在环路”式混合增强智能[57]。这种思路符合 HCAI 理念，即将人的作用引入 AI 系统中，形成以人为中心、融于人机关系的混合智能。例如，在“人在环路”范式中，人始终是 AI 系统的一部分，当系统输出置信度低时，人主动介入调整参数给出合理正确的问题求解，构成提升智能水平的反馈回路[1][48]。另一种方案是在生物学层面上开发“脑在环路”式混增强智能[3]，以生物智能和机器智能深度融合为目标，通过神经连接通道，可以形成对某个功能体的增强、替代和补偿。

第二类混合增强智能是将人类认知模型嵌入 AI 系统中，形成基于认知计算的混合增强智能[1]。从 HCAI 理念分析，这类混合增强智能并不是真正意义上的人机混合增强智能，因为这种系统并非能够保证以人机系统为载体来实现人在人机系统中的中心作用和最终决控权。当然，把人类认知模型引入到机器智能中，对于发展机器智能是非常重要的。

HAII 领域工作将对人机混合增强智能研究和应用发挥重要作用。首先，HAII 领域提倡心理学、认知工程等学科专业人员的合作支持，加速现有心理学等学科成果的转换来支持认知计算的研究，提供有效的认知计算体系架构[49]，例如，为提高对非结构化视听觉感知信息的理解能力和海量异构信息的处理效率，HAII 领域需要支持 AI 界在“感知特征的提取、表达及整合”和“模态信息协同计算”等方面的视听觉信息认知计算研究[50]。HAII 鼓励将基于认知计算方法与“人在环路”方法（系统、生物学层面）整合的工作思路。基于 HCAI 理念，这种思路有助于开发出更强大、可持续发展的、人类可控的 AI。

第二，开展基于 HCAI 理念的人机混合智能控制研究。针对人机混合智能系统控制，目前主要有两种方案：“人在回路控制”和“人机协同控制”[51]。AI 系统在应急状态时人机之间的高效切换是目前重要研究课题。例如，自主武器系统发射后的追踪控制，自主驾驶车应急状态下的高效人机切换。HCAI 理念要求人拥有最终控制权，这需要 AI、人机交互等专业人员的合作，寻找有效解决方案。

第三，开展人机混合增强智能系统的人机交互研究。“人在环路”混合智能系统需要与用户交互的交互设计[30]。不同于传统人机交互，用户交互的对象是 AI 模型，用户界面难以理解，用户与 AI 交互中存在用户意图的不确定性[1]。HAII 研究要求 AI 专业人员与人机交互、人因工程等专业合作，从智能系统、用户、人机交互设计三方面优化系统设计。例如，开发自然式交互设计，选择有效的心理模型。

最后，开展人类高级认知层面上的人机混合增强智能研究。HAII 研究需要 AI、认知神经科学、计算机科学、心理学等专业人员的合作。例如，进一步探索人机融合、脑机融合等方面的研究，今后要在更高的认知层次上为脑机智能的叠加（如学习、记忆）建立更有效的模型和算法[52]； 探索如何将人的决策和经验与机器智能在逻辑推理、演绎推理等方面的优势结合，使人机合作具有高效率[1]。从长远看，人机混合增强智能未来可能形成有效的人机共生[53]，通过个体和群体智能融合等途径，最终在系统和生物学层面上实现人机共生和融合[54]。

### 4.3 人-AI合作

智能技术带来了一种新型人机关系：人-AI 合作，人-AI 系统作为一个组合体比单个实体的工作更加有效[33][35]。人-AI 合作的研究目前在国外是一个热点。HAII 研究需要 AI、心理学和人因工程等学科的合作，从感知、认知、执行三个层面上开展。

在感知层面，为了有效的人-AI 合作，AI 系统需要人的模型来支持系统对人类状态的监控（生理、行为、情绪、愿图、能力等）；AI 系统的人机界面要足够透明，帮助人类了解当前系统状态。 例如，人机之间情景意识（态势感知）分享是人-AI 合作研究的基本问题之一。研究需要了解如何有效实现人-AI 之间基于情景意识模型的的双向沟通[55]，目前还缺少针对人-AI 合作的情景意识模型和测试方法[66]。今后的 HAII 工作需要丰富情景意识理论，为人-AI 合作建模、认知架构、绩效测评提供支持。

在认知层面， PC 时代的人机交互模型已经不能满足智能时代的复杂交互场景。HAII 需要构建符合人-AI 合作的认知和计算模型[57]。人与 AI 之间的互信影响人-AI 合作的绩效，HAII 需要研究信任测量、建模、修复、校正等方面的工作，以及如何量化不同操作场景中人机之间动态化功能交换时所需的信任。不同于传统人机交互，人与 AI 均需要彼此感知并识别交方的意图与情感，今后研究要进一步探索心理模型、意图识别、情感交互等模型，以及在系统设计中如何实现和验证这些模型。

在执行层面，有效的人-AI 合作应该允许在任务、功能、系统等层面上实现决控权在人与 AI 代理之间的分享。决控权的转移取决于人机双向信任、情景意识共享、合作程度等因素。例如，在自动驾驶车领域，HAII 工作需要研究人机控制分享范

式、人机共驾所需的情景意识分享、人机互信、风险评估等，保证车辆控制权在人机之间的快速有效切换，确保人拥有最终控制权（包括远程控制等）[58]。HAII 研究需要了解在什么条件下人机之间如何完成有效切换，是否可以借助人-AI 合作的思路，通过有效人机交互，提供有效的人机控制权转移。 HAII 今后的工作还应该在以下几方面开展。

首先，HAII 需要为人-AI 合作的研究开发新理论、模型以及评估和预测人-AI 合作团队绩效的方法，这些都是传统人机交互中没有遇到的新问题。HAII 领域的工作要支持 AI 建模以及对建模数据的需求（例如，情景意识，行为，意图，信任）[13]，合作开发人-AI 合作在各种应用领域的解决方案。

其次，HAII 领域需要从行为科学等角度、社会层面上来研究人-AI 合作。要研究社会因素（社会责任，道德等）对人-AI 合作的影响，研究如何让 AI 代理担当团队角色并且与人类队友合作[13]； 研究如何从系统设计角度，通过合适的人机交互方式来发展良好的人机关系（信任，情感等）； 研究如何有效支持长期人-AI 合作（如医用机器人）。

第三，HAII 研究要构建人-AI 合作场景所需的人机交互建模。与 AI 人员合作研究对人-AI 合作中人机交互建模构成挑战的理论问题，例如，分布式认知理论，基于上下文的知识表征和知识图谱[59]，人-AI 合作中情感交互、社交互动等方面的认知建模。

最后，HAII 需要在真实的操作、社交环境中研究人-AI 合作[60]。例如，实验室研究表明，与简单机器人的交互可以增强人的协调性，并且机器人可以直接与人合作[61]；人-AI 合作中 AI 与人的认知风格、人格特性等特性相适应时，可增强人机互信与可靠性。今后要在真实社会环境中验证这些人-AI 社会互动，这将有助于优化人-AI 合作的设计。

### 4.4 可解释AI

深度学习等方法会产生 AI“黑匣子”效应，导致用户对 AI 系统的决策产生疑问，该效应可在各类使用中发生，包括 AI 在金融股票、医疗诊断、安全检测、法律判决等领域，导致系统决策效率降低、伦理道德等问题，影响公众对 AI 的信任度。

寻求可解释 AI（eXplainable AI, XAI)已成为 AI 界的一个研究热点，具有代表性的是 DARPA[62]在 2016 年启动的项目。该项目集中在：(1) 开发或改进 ML 技术来获取可解释算法模型；(2) 借助于先进的人机交互技术，开发可解释 AI 的用户界面模型；(3) 评估心理学解释理论来协助何解释 AI 的研发。

经过多年的努力，AI 界开始认识到非 AI 学科在可解释 AI 研究中的重要性[63]。Miller[64]的调查表明，大多数可解释 AI 项目仅在 AI 学科内展开。 许多 AI 人员采用“以算法为中心”的方法，加剧了算法的不透明[65]。一些 AI 人员没有遵循 HCAI 理念，通常为自己而不是用户构建可解释 AI [66]。研究还表明，如果采用行为科学方法，侧重于用户而不是技术，针对可解释 AI 的研究将更有效[64]。

今后 HAII 领域的工作主要有以下几方面。首先，研究和开发“以人为中心的可解释 AI”解决方案。HAII 领域要从人机交互、心理学、人因工程等方面来寻找解决方案。以往许多研究是基于静态和单向信息传达式的解释，今后 HAII 工作要研究探索式、自然式、交互式解释来设计可解释界面[67-68]。

第二，HAII 领域提倡可解释 AI 研究要进一步挖掘心理学等模型。尽管这些理论和模型通常是基于实验室研究产生，可解释 AI 研究应该善于利用这些模型，同时验证它们的可行性[68]。HAII 的工作可以利用本身交叉学科的特点起到一个中间桥梁作用，加快理论转换，构建有效的界面或计算模型。

最后，HAII 领域要开展可理解 AI 的研究和应用。可解释 AI 也应该是可理解的[68]，可解释性是必要条件，但不是充分条件。从 HCAI 理念考虑，可理解 AI 方案应满足终端用户的需求(例如知识水平)。这方面的研究需要行为科学方法的支持以及实验验证[69]。

### 4.5 人类可控自主化

智能自主化技术正在走进人们的工作和生活，但是已有人开始混淆自动化与自主化的概念，这可能导致对技术不恰当的期望和误用[10]。自动化技术按照固定算法、逻辑和规则而产生确定的机器行为。智能系统会拥有不同程度的类似于人的智能（自适应、自我执行等能力），系统输出具不确定性，有可能出现偏差的机器行为。AI 的自主化特征对安全和大众心理等负面影响还没有引起足够的重视[70]。

从 HCAI 理念出发，我们提倡基于“人类可控 AI”设计目标的人类可控自主化设计，即智能自主化系统需要人类的监控，人类操作员具有最终的决控权（通过直接或远程操控等方式）[4]。

人因工程界已经对一些复杂领域（航空、航天等）中的自动化系统开展了广泛的研究，已达成共识[71-72]。许多复杂自动化系统存在脆弱性，在设计所规定的操作场景中运行良好，但是在遇到意外事件时，可能导致操作员的“自动化惊讶”现象[71]：操作员无法理解自动化正在做什么，为什么这样做。

Endsley[70]认为在智能自主化系统中，随着系

统自主化程度的提高，各单项功能的“自动化”水平也相应提高，这可能会导致操作员对这些功能的关注度降低，在应急状态下出现“人在环外”的现象。对近几年发生的多起自动驾驶车致命事故调查表明，界面模式混淆、“人在环外”、过度信任等问题正是以往自动化系统中出现的典型问题[73-74]。

智能技术中潜在的自主操作性等特征也会造成人们对该技术产生类似于对自动化的过度信任。具有学习能力的自主化系统意味着其操作结果的不确定性可能以意想不到的方式发展，有可能给操作员带来比自动化更强烈的“自动化惊讶”体验。

今后 HAII 领域的工作可从以下几方面考虑：

首先，HAII 领域要针对一些自主化的基本问题展开研究。从人机交互角度充分理解 AI 自主化特性对人机交互设计的影响，研究自主化对操作员期望、角色等的影响，研究自主化对操作员情绪应激、认知能力、人格特质和沟通属性的影响[75-76]。

其次，HAII 领域要在自主系统开发中实现 HCAI 理念的“人类可控 AI”设计目标。目前，尽管人机交互等专业人员参与了自主化系统（如自动驾驶汽车）的研发，但是频频发生的事故提醒我们评估目前的方法[11]。SAE[77]认为 L4-L5 等级的自动驾驶车不需要人类监控和干预，我们质疑 SAE 忽略了自动化和自主化之间的本质差异[78]，可能对设计、安全、标准化和认证等产生不利影响。高等级自动驾驶汽车是一个“移动式”自主化系统，不是传统的自动化系统。基于 HCAI 理念，要从人-AI 合作、人机互信、态势感知共享、自主化共享等角度探索自主化设计，实现有效的人机共驾及交接[79]，任何等级的自动驾驶车都需要确保人是系统的最终决控者（包括远程控制方式）。

最后，实现针对自主化系统的“有意义的人类控制”（meaningful human control）设计目标[46]。HAII 工作要落实该目标的实现：（1）通过“人在环路”、人机交互设计，保证应急状态下人类可接管或中断系统运行；（2）在重要的自主化系统中安装“故障追踪系统”来实现设计改善和人机故障问责制[46]，推动 HCAI 理念中“人类可控 AI”设计目标的实现。

### 4.6 智能人机交互

智能化人机交互为 HAII 领域带来了挑战和机遇。AI 系统丰富的应用场景和用户需求需要有效的人机交互范式[80]。现有人机交互方式（如 WIMP）局限于有限的感知通道、交互带宽不足、输入/输出带宽不平衡、交互方式不自然等问题，已有研究提出了 Post-WIMP 的范式[81]，这些范式需要 HAII 工作的验证。多模态融合及并行交互范式是今后 HAII 的重要研究内容，这方面研究目前主要在 AI、计算技术界展开，HAII 应该提供跨学科支持。HAI 还要在情境感知、意图理解等方面取得更大突破[81-83]。

人-AI 合作的新型人机关系对人机界面设计提出了新要求。传统人机界面主要基于“刺激-反应”理念的“指令顺序”式交互，针对智能人机交互（情感、意图识别、上下文检测等）的多模态余度式交互，HAII 要设计有效的人机界面来支持人-AI 合作所需的情景意识分享、人机互信、人机控制分享等。

HAII 领域要开发针对 AI 系统的人机交互设计标准。现有的标准主要是针对非 AI 系统，国际标准化组织正在起草人-AI 系统交互的设计标准（ISO 9241-810）[42]，已有一些针对 AI 系统的人机交互设计指南[17][64]，但是还需要 HAII 领域的贡献。

### 4.7 伦理化AI设计

HCAI 理念推崇的伦理化 AI 设计目标是 AI 界目前普遍关心的重要问题。研究表明，AI 人员在职业培训中通常缺乏应用伦理规范进行设计的培训，AI 界已经认识到伦理化 AI 设计需要多学科的合作[84]。专业组织和企业已发布了多套 AI 伦理准则，但是研究表明在 AI 系统开发中如何有效落实这些规范有待进一步努力，一些专业人员是在开发后期而不是过程中考虑伦理化设计[85]。因此，HAII 领域的一项重要工作是将伦理化 AI 设计落实在开发过程中。

首先，HAII 领域要开展针对 AI 机器行为的研究。2019 年 MIT 等大学的多名学者在《自然》上发文建议开展 AI 机器行为的研究[9]。目前从事机器行为研究主要是 AI 人员，没有受过行为科学的训练。AI 机器行为的非确定性，需要从算法、数据、培训、测试等方面来研究影响因素，避免算法偏差。目前，已有基于 HCAI 理念的机器行为研究，例如，“以人为中心的机器学习”、交互式机器学习等方法[29][66]。这些方法有助于在开发中解决 AI 系统极端行为、公平性等问题。

第二，HAII 领域可采用跨学科的方法论来支持伦理化 AI 设计，将人机交互所倡导的迭代式设计和测评方法应用在模型算法训练中，收集算法培训数据，定义用户预期结果并且转化成有效的输入数据，利用早期原型开展用户体验测评，通过迭代式设计和测试来减小算法偏差[86]。

第三，采用“有意义的人类控制”方法将伦理化 AI 落实在系统设计中[46][70]。系统设计要保证：（1）操作员能够对所使用的自主化技术做出知情且有意识的决策；（2）操作员有足够的信息来确保

采取合法的行动。另外，在系统设计、测试、专业人员培训等方面采取措施来确保人类对系统的有效控制。“有意义的人类控制”与 HCAI 理念一致，有利于在设计中实现伦理化 AI 的目标。

最后，HAII 领域要优化 AI 的开发实践来支持伦理化 AI 设计。研究表明，目前 AI 界缺乏有效的伦理化 AI 设计方法论、指导设计选项的技术细节或详细示例[85]。HAII 领域可从“人在环路”AI、人-AI 合作等设计思路方面提供帮助。如何将伦理化 AI 原则嵌入到开发流程、如何提高伦理化 AI 准则对 AI 工程师行为的影响、如何提升 AI 工程师的伦理化设计技能等方面的问题都需要 HAII 的多学科方案。

## 5 HAII 研究和应用的挑战及对策

作为一个新兴领域，在实现 HCAI 理念的实践中，HAII 研究和应用必定面对挑战。

第一，AI 人员缺乏对 HCAI 理念的理解。许多 AI 人员在开发中基本按照“以技术为中心”的理念，一些人员认为人机交互无法解决的问题目前已被 AI 技术解决（如语音输入），人机界面不必考虑用户体验；而人机交互等专业人员往往在 AI 项目产品需求定义后才参与项目，限制了他们对 AI 系统设计的影响，并且导致一些 AI 项目的失败[87-89]。

第二，AI 系统研发中缺乏有效的跨学科合作。AI 人员与非 AI 人员之间缺乏有效的沟通语言，非 AI 人士缺乏必要的 AI 知识，而 AI 人士缺乏对其他学科的了解[90]。目前，已有研究提出了基于 HCAI 理念的 AI 与人机交互专业人员的“配对式 AI 合作开发流程”[46]以及提升的人机交互方法[91]。

第三，跨学科合作缺乏有效的方法。一些人机交互人员在 AI 系统开发中仍然采用传统针对非 AI 系统的方法；许多 AI 人员不易接受其他学科的方法。目前，人们已经提出了一些方法，例如，“AI 优先”方法、“AI 作为设计材料”[43][45]。

为有效实现 HCAI 理念和开展 HAII 工作，我们从以下 3 个层面提出建议（如图 3 所示）。

首先，在 AI 研发团队层面，建立多学科团队和采用跨学科方法。AI 带来的新问题只有采用多学科合作才能找到有效方案。基于 HCAI 理念来优化现有的 AI 研发流程，例如，在开发初期制定 HCAI 设计目标，优化在各个开发流程节点上协同合作。

其次，在 AI 研发企业组织层面，培育基于 HCAI 理念的组织文化，制定基于 HCAI 的开发标准指南，提供 HCAI 研发资源（跨学科人力资源、HAII 项目、跨学科研究设备等），建立高效的 AI 研发团队。

最后，在 AI 研发社会层面，培养具备 HCAI 理念的跨学科复合型人才。例如，在高校开设“AI 主修 + 辅修”、“主修 + AI 辅修”本科专业，培养针对 HAII 关键问题的硕博生；制定 AI 发展策略、法规等；开展跨行业、跨学科攻关项目，提倡学术界和工业界之间的协作；设立政府专项基金来支持 HAII 项目；建立完善的产学研相结合的科研体系，在一些关键行业开展 HAII 研究和应用。

## 6 结论

一、当前的智能时代呈现出“技术提升 + 应用开发 + 以人为中心”的阶段特征。拥有一定程度的学习、自适应、独立执行等能力是智能技术的独特自主化特征，赋予了机器在人机系统中新的角色，带来了新型人机关系：人-AI 合作。与人-计算机交互相比，人-AI 交互带来了许多新特征和新问题，给 AI 系统研发带来了新挑战，促使我们重新评估目前所采用的、基于非 AI 系统的研发策略。

二、为有效解决 AI 系统研发中的新挑战，本文阐述了 HCAI 的理念，它强调三方面工作的有机结合。(1)“技术”：提倡 AI 技术的发展策略应该注重人类与机器智能的优势互补，将人的作用引入 AI 系统中，从而产生更强大、可持续发展的人机混合增强智能。AI 的目的是提升人的能力，不是取代人；(2)“人”：强调从人的需求出发，为人类提供有用、可用、人类拥有最终决控权的 AI 系统；(3)“伦理”：从人类伦理出发，AI 系统研发要避免产生伦理道德等问题，目的是开发出伦理化、负责任的 AI。

三、为实现 HCAI 理念在 AI 研发中的落实，本文系统提出 HAII 这一跨学科新领域。HAII 领域可以为各学科提供一个有效的合作平台，注重 AI 技术的新挑战和新问题，有助于推动 HCAI 理念在 AI 研发中的落实，开发出安全、可靠、可信赖的 AI 系统。

四、针对 AI 带来的新问题，我们提出今后 HAII 研究和应用的重点方向，它们包括人类智能增强技术、人机混合增强智能、人-AI 合作、可解释 AI、人类可控自主化、智能人机交互、伦理化 AI 设计。

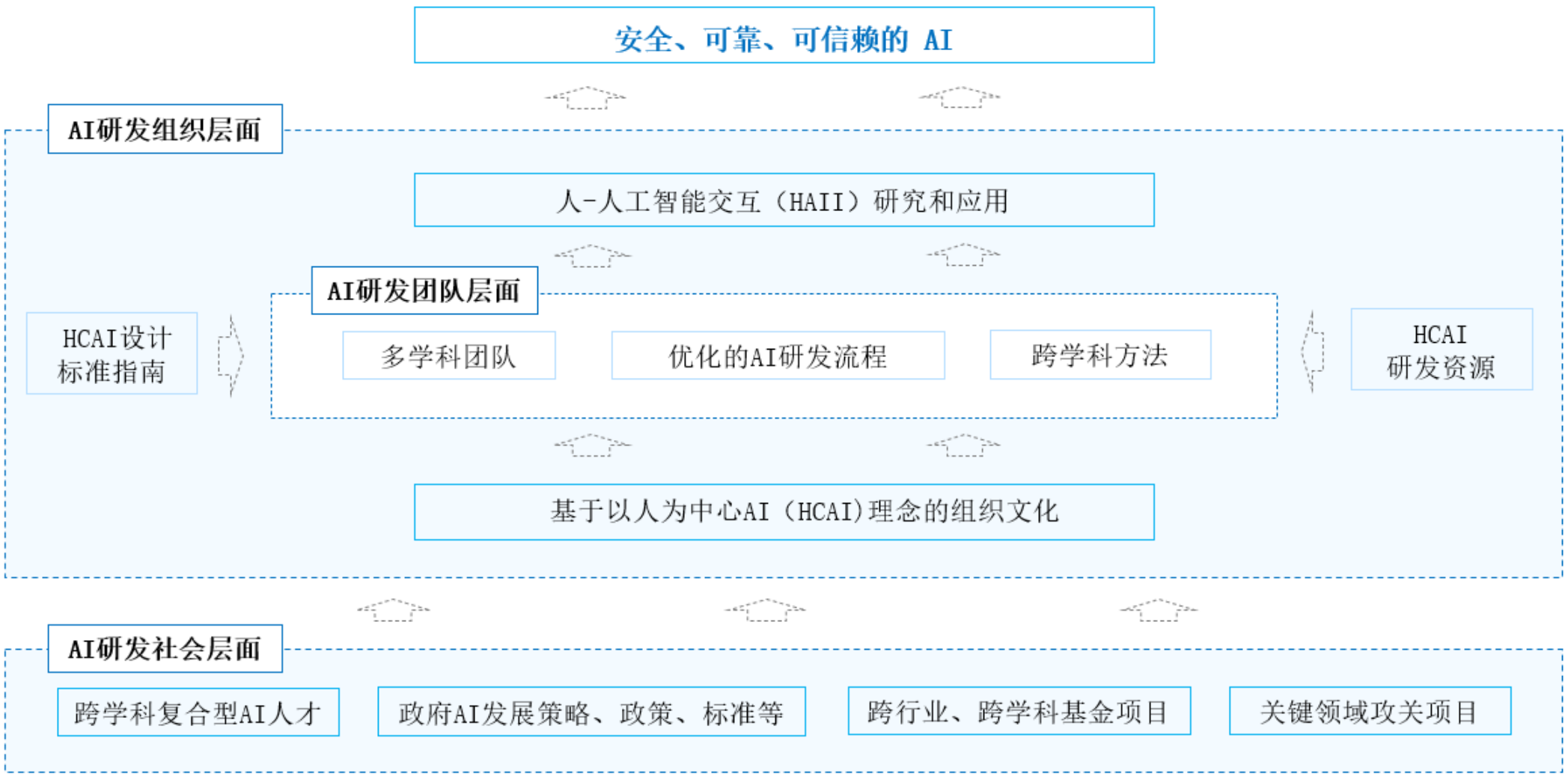

图 3 有效实现HCAI理念以及开展HAII研究和应用的“三层面”策略

Figure 3 The “3-layer” strategy for effectively implementing HCAI and carrying out HAII research and application

五、针对今后 HAII 实践面对的挑战，从研发团队、研发组织、研发社会环境 3 个层面上，我们建议：设立多学科研发团队，采用跨学科方法，优化现有 AI 研发流程；培育 HCAI 理念的组织文化，开发 HCAI 标准指南和研发资源；培养跨学科复合型 AI 人才，政府出台相关政策和法规，积极开展跨学科、跨行业以及关键行业领域的 HAII 研究和应用。我们展望只要各学科和各行业一起努力，一个基于“以人为中心 AI”理念的智能社会时代将会到来。

# 参考文献

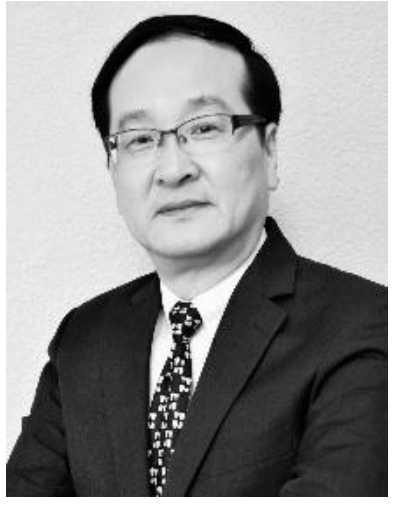

许为，留美心理学博士和计算机科学硕士，教授；国际知名 IT 企业资深研究员、人机交互技术委员会主席。任国际标准化组织（ISO）人-系统交互专业委员会成员、International Journal of Human-Computer Interaction 等 SCI 期刊编委、中国认知工效学分会和工程心理学分会理事。30 多年来一直在国内外高校、知名 IT 和民用航空企业从事人机交互和人因工程方面的研究、设计以及标准开发；主持或参与许多国家省部级项目，获众多研究和设计奖项，成果应用在国内外多种飞机型号和 IT 产品；出版中英文书著 5 部，在国内外人机交互、心理学、航空、计算技术、人因工程等核心学术期刊发表大量论文；主持或参与开发国内外人因工程、人机交互设计标准 20 余部。

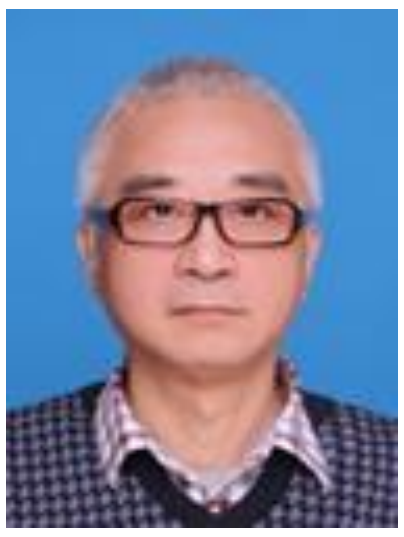

葛列众，博士，教授，中国心理学会工程心理学专业委员会现任主任委员。主要研究方向有人机交互、面孔认知、用户可用性和用户体验研究。曾主持国家省部级项目 8 项；主持华为等公司横向课题 30 余项；主持国际合作项目 4 项。在国内外学术刊物发表论文 174 篇。2019 年获得中国心理学会"学科建设成就奖"，2019 年领衔团队获得中央军委装备发展部等 5 部委颁发的"中国航天载人工程突出成就集体奖"。

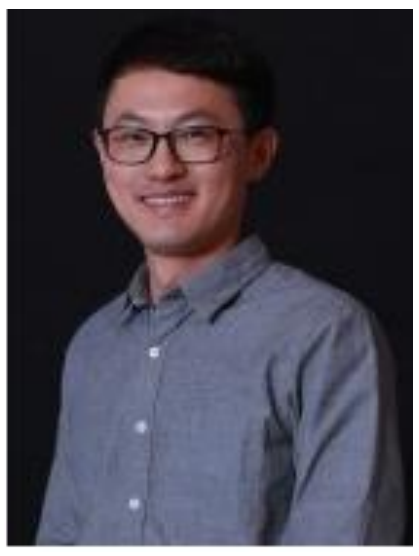

高在峰，博士，教授，浙江大学心理与行为科学系副主任，长江学者青年学者，中国认知工效学会理事，《应用心理学》杂志编委。主要从事认知心理学、工程心理学的相关研究，主持国家省部级项目 9 项；以第一作者或通讯作者在 Psychological Science， Cognition，International Journal of Human-Computer Interaction 等发表 SCI/SSCI 论文 40 余篇，担任 Automotive Innovation，Cognition，Emotion， Cortex 等 40 余个国际核心学术期刊的审稿人。